# EQUITY IN STARTUPS

**An analysis of the equity split in more than 400 successful startups.**

**September 2017**

**HERVÉ LEBRET**

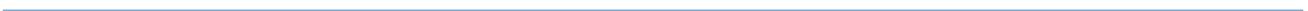

I describe in this report equity allocation in startups based on more than 400 real cases, most of them being companies which went public or had filed to go public. Such companies disclose a lot of information in the documents they must provide before their initial public offering (IPO) and in particular they describe the equity structure at the time of that event and how it evolved from the company incorporation onwards. I could compile and analyze that rich information with a focus on the allocation of equity to founders, employees and managers, investors, board members. The large sample makes it possible to also study that allocation relatively to the fields of activity, periods of foundation, and the geographic origin of startups.

## A Silicon Valley Phenomenon

Equity has been a major component of value creation in startups, particularly in Silicon Valley. The topic is still very much debated with regular ebb and flow cycles about the merits of stock options, ordinary and preferred shares, and many other equity mechanisms. Interestingly enough, the current status of equity allocation in startups is the output of a slow process which began in the early days of Silicon Valley in 1957, with Fairchild Semiconductor, maybe the first startup ever [1]: *"Fairchild Camera and Instrument put up the million and a half dollars that we [founders] felt we needed in return for which they got an option to buy all of our stock. And the stock was divided [equally] for each of the eight [founders]. [... Fairchild] had a very, what I call Eastern mentality in that they didn't want anybody to have any options in stock and the eight entrepreneurs who started Fairchild Semiconductor decided individually and together that they would gradually peel off, and form their own enterprises because they couldn't get any more equity in, and a lot of the people there felt that they should be giving equity to some of the people who hadn't helped start the company but were instrumental in its success. And Fairchild Camera and Instrument was unwilling to do that. So gradually they peeled off and finally by 1968 there were only Noyce and Moore left."* This anecdote shows that a lot of frustration was created because of bad stock allocation and in a way, indirectly but positively, that frustration created incentives for more startup foundations and new equity mechanisms.

Silicon Valley remains today the center place for startup creation and after sixty years, equity allocation has been "optimized" to align the interests of all startup stakeholders, that is founders, investors and employees mainly, but also the general public when a startup goes public and licensing entities such as universities when a startup is created with the intellectual property (IP) of such entities. Although all this is quite well known [2, 3], I noticed there was still room for giving more information about equity allocation in startups.

Despite what was claimed above, there is not one single or best way to allocate equity in startups. Don Valentine, the founder of Sequoia Capital, one of the most famous venture capital firms in Silicon Valley, which invested in firms including Apple, Google, Oracle, PayPal, YouTube, Instagram, Yahoo! and WhatsApp describes it his way [4]: *"When people come as a team (usually it is three or four people and typically heavyweight on engineering), it is a complex process. But I think all of us have seen it in the earlier days, times when I can remember saying, "Well, look, we'll put up all the money, you put up all the blood, sweat and tears and we'll split the company", this*

*with the founders. Then if we have to hire more people, we'll all come down evenly; it will be kind of a 50/50 arrangement. Well, as this bubble got bigger and bigger, you know, they were coming and saying, "Well, you know, we'll give you, for all the money, 5 percent, 10 percent of the deal." And, you know, that it's a supply and demand thing. It's gone back the other way now. But, in starting with a team, it's a typical thing to say, well, somewhere 40 to 60 percent, to divide it now. If they've got the best thing since sliced bread and you think they have it and they think they have it, you know, then you'll probably lose the deal because one of these guys will grab it."*

## Equity Allocation in Startups: the General Idea

Equity allocation in startups begins with founders. As it happened with Fairchild or as Valentine describes it, startup founders are usually a team of engineers. They can be 2, 3, 4 or more founders. There are also cases of single-founder startups. The founding team splits the initial shares in the company. Sometimes equally, sometimes not. Sometimes, the founders also need IP from external entities, usually their former employer, and in many cases the licensing entity receives founders' equity in exchange for the IP license.

Once the founders have agreed on the initial founding split, they may or may not allocate new shares. This has become a standard way for incentivizing startup partners in Silicon Valley. Less in other regions, even if this becomes common practice. These partners are mostly investors and employees. Investors receive *preferred shares* in exchange for the capital they invest in startups, whereas founders usually own *ordinary shares*. Employees receive *stock options* for their "sweat and tears", which may be converted in ordinary shares at some point in the startup life [3, 5].

It may be surprising for some readers to learn that there is no real equality between all stakeholders. Preferred shares have more privileges than ordinary or common shares and stock options are not even real shares. But usually, at a liquidity event such as an IPO or the sale of the company, all shares are converted into common shares. Then there is more equality. Even then, the voting power of shares maybe different… Business and democracy do not often go along well!

## Equity Allocation in more than 400 Successful Startups

For many years, I have been using Nesheim's template [3] to build capitalization tables very similar to the one shown in table i in the appendix. At the date of writing this report, I had compiled more than 400 cases [6]. The next tables summarize my results. First, tables 1 to 3 describe what these startups are about in terms of fields of activity, periods of foundation and geography with figures on their financial situation such as amounts of venture capital (VC) raised, yearly sales and profits before the IPO as well as their number of employees and number of years to go public.

The following tables (4 to 6) give the equity allocation, i.e. how much the group of founders keep at IPO, how much employees and investors get as well as the allocation to the general public at the IPO. An additional interesting element is added, that is the average age of the group of founders. Again these elements are given relatively to fields of activities (table 4), periods of foundations (table 5) and geography (table 6).

| Field | # Startups | Years to IPO | VC ($M) | 1st round ($M) | Sales ($M) | Income ($M) | Employees |
|---|---|---|---|---|---|---|---|
| Biotech | 114 | 8.5 | 90 | 11 | 11 | -17 | 70 |
| Medtech | 20 | 11.1 | 91 | 4 | 23 | -13 | 181 |
| Internet | 96 | 6.5 | 281 | 6 | 283 | 26 | 1215 |
| Software | 56 | 9.5 | 75 | 6 | 110 | 1 | 586 |
| HW/Comp./Tel | 60 | 7.9 | 101 | 7 | 91 | -14 | 414 |
| Semiconductor | 35 | 8.0 | 54 | 6 | 60 | -6 | 376 |
| Energy/Env. | 16 | 6.6 | 208 | 5 | 50 | -41 | 425 |
| Other | 6 | 8.0 | 137 | 6 | 169 | -15 | 391 |
| Overall | 403 | 8 | 138 | 8 | 109 | -4 | 519 |

*Table 1: Data on startups vs. fields of activity*

| Period of foundation | # Startups | Years to IPO | VC ($M) | 1st round ($M) | Sales ($M) | Income ($M) | Employees |
|---|---|---|---|---|---|---|---|
| 1965 | 2 | 3 | 4 | 2 | 4 | -1 | 342 |
| 1970 | 1 | 3 | 6 | 1 | 7 | 1 | 137 |
| 1975 | 5 | 7 | 6 | 1 | 60 | 8 | 529 |
| 1980 | 15 | 5 | 22 | 3 | 29 | 1 | 301 |
| 1985 | 10 | 12 | 41 | 12 | 82 | -7 | 835 |
| 1990 | 43 | 10 | 47 | 4 | 38 | -5 | 255 |
| 1995 | 98 | 8 | 91 | 5 | 150 | 27 | 628 |
| 2000 | 114 | 9 | 109 | 11 | 132 | 4 | 409 |
| 2005 | 97 | 7 | 190 | 8 | 104 | -37 | 650 |
| 2010 | 18 | 3 | 704 | 10 | 46 | -50 | 617 |
| Overall | 403 | 8 | 138 | 8 | 109 | -4 | 519 |

*Table 2: Data on startups vs. periods of foundation*

| Geography | # Startups | Years to IPO | VC ($M) | 1st round ($M) | Sales ($M) | Income ($M) | Employees |
|---|---|---|---|---|---|---|---|
| Silicon Valley | 155 | 7 | 212 | 6 | 109 | -19 | 447 |
| Boston Area | 56 | 8 | 93 | 7 | 47 | -12 | 282 |
| California | 33 | 10 | 89 | 7 | 23 | -16 | 202 |
| West Coast | 17 | 8 | 79 | 7 | 130 | -23 | 574 |
| East Coast | 34 | 9 | 96 | 10 | 32 | -17 | 144 |
| Midwest | 17 | 6 | 171 | 7 | 114 | -34 | 731 |
| France | 28 | 10 | 38 | 4 | 63 | -5 | 445 |
| Switzerland | 17 | 8 | 75 | 22 | 21 | -12 | 122 |
| UK | 9 | 10 | 88 | 6 | 382 | 84 | 709 |
| Other EU | 13 | 9 | 82 | 4 | 188 | -5 | 813 |
| Israel | 4 | 11 | 180 | 4 | 78 | 5 | 294 |
| China | 12 | 8 | 122 | 10 | 771 | 308 | 4086 |
| Canada | 1 | 14 | 43 | 5 | 13 | 0 | 183 |
| Other | 7 | 9 | 45 | 19 | 107 | -1 | 992 |
| Overall | 403 | 8 | 138 | 8 | 109 | -4 | 519 |

*Table 3: Data on startups vs. geography*

| Field | Founders' age | Founders | Employees | Investors | IPO Shares |
|---|---|---|---|---|---|
| Biotech | 45 | 7% | 15% | 57% | 23% |
| Medtech | 42 | 8% | 17% | 54% | 22% |
| Internet | 34 | 17% | 22% | 48% | 15% |
| Software | 34 | 17% | 27% | 42% | 16% |
| HW/Comp./Tel | 37 | 12% | 27% | 47% | 15% |
| Semiconductor | 38 | 13% | 26% | 44% | 18% |
| Energy/Env. | 38 | 8% | 19% | 56% | 18% |
| Other | 39 | 10% | 21% | 52% | 20% |
| Overall | 38 | 12% | 22% | 50% | 18% |

*Table 4: Data on equity split vs. fields of activity*

| Period of foundation | Founders' age | Founders | Employees | Investors | IPO Shares |
|---|---|---|---|---|---|
| 1965 | 37 | 25% | 23% | 36% | 16% |
| 1970 | 32 | 8% | 22% | 52% | 18% |
| 1975 | 29 | 32% | 34% | 22% | 12% |
| 1980 | 35 | 15% | 28% | 39% | 19% |
| 1985 | 36 | 15% | 28% | 48% | 17% |
| 1990 | 35 | 18% | 24% | 39% | 20% |
| 1995 | 38 | 11% | 23% | 50% | 17% |
| 2000 | 38 | 10% | 21% | 52% | 18% |
| 2005 | 40 | 10% | 19% | 54% | 19% |
| 2010 | 41 | 19% | 14% | 54% | 21% |
| Overall | 38 | 12% | 22% | 50% | 18% |

*Table 5: Data on equity split vs. periods of foundation*

| Geography | Founders' age | Founders | Employees | Investors | IPO Shares |
|---|---|---|---|---|---|
| Silicon Valley | 37 | 13% | 27% | 46% | 15% |
| Boston Area | 42 | 9% | 19% | 51% | 22% |
| California | 43 | 7% | 19% | 56% | 18% |
| West Coast | 35 | 19% | 22% | 47% | 14% |
| East Coast | 41 | 9% | 19% | 54% | 20% |
| Midwest | 42 | 9% | 17% | 55% | 20% |
| France | 35 | 17% | 13% | 51% | 24% |
| Switzerland | 40 | 14% | 14% | 55% | 21% |
| UK | 35 | 8% | 24% | 48% | 21% |
| Other EU | 31 | 18% | 16% | 55% | 26% |
| Israel | 33 | 14% | 17% | 55% | 14% |
| China | 32 | 22% | 20% | 43% | 15% |
| Canada | 24 | 23% | 39% | 17% | 21% |
| Other | 35 | 15% | 19% | 49% | 19% |
| Overall | 38 | 12% | 22% | 50% | 18% |

*Table 6: Data on equity split vs. geography*

# Main Facts about the 400 Successful Startups

Let me summarize again the main facts gathered from these companies. A simple look at data shows that at IPO (or exit) founders keep around 10% of their company whereas investors own 50% and employees 20%. The remaining 20% goes to the general public at IPO. Valentine is wrong when he claimed founders and investors would split equally! Of course, this is a little too simplistic. For examples founders keep more in Software and Internet startups and less in Biotech and Medtech. There could be a lot more to add but I let the reader focus on what possibly interests her. Additional interesting points are:
- The average *age of founders* is 38 but higher in Biotech and Medtech and lower in Software and Internet.
- It takes on average 8 years to go public after raising a total of $138M, including a first round of $8M in *VC money*.
- On average, companies have about $110M in sales and are slightly profitable, with 500 employees at IPO time. But again there are differences between Software and Internet startups which have more *sales* and *employees* and positive *income* and Biotech and Medtech startups which have much lower revenue and headcount and negative profit.

The attentive reader might have noticed that summing the percentages in the horizontal lines in tables 4-6 does not make 100% but usually more! No mistake here but a point of caution is necessary: when a company did not go public, the IPO percentage was not put to 0% but not considered. The same rule was used for the other dimensions. The average values take into account only companies which had such data and did not take the others into account, which explains the discrepancy.

## Allocation of Shares to Managers

Tables ii to iv in appendix further describe the equity allocation to non-founder managers. These are first the Chief Executive (CEO), then all officers reporting to him, that is Vice-Presidents (VP) and other Chief Officers (CXO), in particular the important Chief Financial (CFO). Finally independent board members (directors) are also critical individuals helping in the company strategy and they are often remunerated with startup shares. Again the reader may focus on fields, periods or geographies she is interested in but let me summarize here the results:
- The *CEO* owns about 3% of the startup at exit. This is 4x less the founding group and depending when she (although it is too often a "he") joined it would mean up to 20% close to foundation (assuming the founders would keep 80% and allocate the delta to the CEO)
- *CEOs* are non-founders in about 36% of the cases, more in biotech (42%) and Medtech (35%) than Internet (31%) and Software (25%), more in Boston (48%) than Silicon Valley (43%).
- The *Vice-Presidents and Chief Officers* own about 1% and the Chief Financial around 0.6%.
- Finally, an *independent director* gets about 0.3% of the equity at IPO. If we consider again that the founders are diluted by a factor 8x from their initial 100% to about 12%, it means a director should have about 2-3% if he joins at inception.

## IP Licensing and Equity

Intellectual property licensing is not an easy element of information to obtain. Most IPO documents do not disclose the numbers when there is a license of IP from a university or a corporation. Still, I could get some figures. About 10% of the sample (38 startups) disclosed some information about IP licenses. Table v in appendix gives the exhaustive information.

The results are consistent with what is disclosed by universities [7]. In the past universities owned about 10% of a startup at creation in exchange for an exclusive license on IP. More recently, this has been more 5% non-diluted until significant funding (Series A round). Table v indeed shows on average such percentages with an average $4.5M $1^{st}$ round funding. At IPO, this represents about 1% equity stake. The reader should be aware that licenses also include royalty payments on sales, but this is a less accepted scheme in Silicon Valley and also outside of Biotech.

## Conclusion

All the data gathered is quite consistent with the literature I know as well as with my practice around startups as a former venture capitalist as well as an academic professional. This remains however a heated debate, particularly in Europe and probably outside of the US technology clusters. Founders do not like dilution; they are not always convinced by the value of stock-options and do not like preferred shares. But the fact is that investors own about half a startup at exit and the typical 20% ESOP seems to be backed by history and statistics. Founders do not always like giving equity to universities either. Again, this is not a rule, just common practice in Silicon Valley. I hope this report will be helpful and comments are more than welcome…

# Appendix

# A Capitalization Table Example

|   | Activity | High-Tech |   | Company | Computer Systems, CS |   | Incorporation |   |
|---|---|---|---|---|---|---|---|---|
|   | Town, St | Stanford, CA |   | IPO date | 5-Jun-06 |   | State | DE |
| f= | founder | Price per share $20 |   | Market cap. | $911'792'206 |   | Date | Sep-02 |
| D= | director | Symbol |   | URL |   |   | years to IPO | 3.7 |

|   | Title | Name | Ownership | | | | | Number of shares/stock | | | | | Value |
|---|---|---|---|---|---|---|---|---|---|---|---|---|---|
|   |   |   | Founder's | Series A | Series B | PreIPO /C | Post IPO | Founder's | Series A | Series B | PreIPO /C | Post IPO |   |
| f | CTO | PhD | 49.0% | 19.6% | 13.5% | 11.8% | 10.7% | 4'900'000 | 4'900'000 | 4'900'000 | 4'900'000 | 4'900'000 | $98'000'000 |
| f | Chief Scientist | Professor | 16.0% | 6.4% | 4.4% | 3.9% | 3.5% | 1'600'000 | 1'600'000 | 1'600'000 | 1'600'000 | 1'600'000 | $32'000'000 |
| f | VP Bus. Dev. | Biz | 31.0% | 12.4% | 8.5% | 7.5% | 6.8% | 3'100'000 | 3'100'000 | 3'100'000 | 3'100'000 | 3'100'000 | $62'000'000 |
|   | CEO |   |   |   | 8.3% | 7.2% | 6.6% |   |   | 3'000'000 | 3'000'000 | 3'000'000 | $60'000'000 |
|   | VP S&M |   |   |   | 1.9% | 1.7% | 1.5% |   |   | 700'000 | 700'000 | 700'000 | $14'000'000 |
|   | VP Eng. |   |   |   |   | 1.0% | 0.9% |   |   |   | 400'000 | 400'000 | $8'000'000 |
|   | VP Prods |   |   |   |   | 1.0% | 0.9% |   |   |   | 400'000 | 400'000 | $8'000'000 |
|   | CFO |   |   |   |   | 0.5% | 0.4% |   |   |   | 200'000 | 200'000 | $4'000'000 |
|   |   |   |   |   |   |   |   | 400'000 | 400'000 | 400'000 | 400'000 | 400'000 |   |
|   | Founders and managers |   | 100.0% | 40.0% | 37.7% | 35.4% | 32.2% | 10'000'000 | 10'000'000 | 13'700'000 | 14'700'000 | 14'700'000 | $294'000'000 |
|   | Other common |   |   |   |   | - | - |   |   |   |   |   |   |
|   | Total common before options |   | 100.0% | 40.0% | 37.7% | 35.4% | 32.2% |   | 10'000'000 | 13'700'000 | 14'700'000 | 14'700'000 | $294'000'000 |
|   | Options-Granted |   |   | 4.0% | 5.7% | 6.0% | 5.5% |   | 1'000'000 | 2'072'727 | 2'500'000 | 2'500'000 | $50'000'000 |
|   | Options-Available |   |   | 16.0% | 4.1% | 2.7% | 3.3% |   | 4'000'000 | 1'500'000 | 1'111'688 | 1'500'000 | $30'000'000 |
|   | Options-Total |   |   | 20.0% | 9.8% | 8.7% | 8.8% |   | 5'000'000 | 3'572'727 | 3'611'688 | 4'000'000 | $80'000'000 |
|   | Sub-total |   |   | 60.0% | 47.5% | 44.1% | 41.0% |   | 15'000'000 | 17'272'727 | 18'311'688 | 18'700'000 | $374'000'000 |
|   | Investors (VCs) |   |   | 40.0% | 52.5% | 46.0% | 41.9% |   | 10'000'000 | 19'090'909 | 19'090'909 | 19'090'909 | $381'818'180 |
|   | Investors (others) |   |   |   |   | 9.9% | 9.0% |   |   |   | 4'098'701 | 4'098'701 | $81'974'026 |
|   | Total- Investors |   |   | 40.0% | 52.5% | 55.9% | 50.9% |   | 10'000'000 | 19'090'909 | 23'189'610 | 23'189'610 | $463'792'206 |
|   | Total - PreIPO |   | 24.1% | 100.0% | 100.0% | 100.0% | 91.9% |   | 25'000'000 | 36'363'636 | 41'501'298 | 41'889'610 | $837'792'206 |
|   | IPO |   |   |   |   |   | 7.7% |   |   |   |   | 3'500'000 | $70'000'000 |
|   | Option (underwriters) |   |   |   |   |   | 0.4% |   |   |   |   | 200'000 | $4'000'000 |
|   | Total outstanding |   | 21.9% |   |   |   | 100.0% |   |   | 36'363'636 | 41'501'298 | 45'589'610 | $911'792'206 |

| Number of employees | 2 | 7 | 25 | 70 | 200 |
|---|---|---|---|---|---|

* The difference between common shares and options is very small. In this case, the number of non-founder shares and ESOP is maintained to 20% of the company at each VC round

| IPO | Total cash before fees | $70'000'000 |
|---|---|---|
|   | Paid to underwriters | $4'900'000 |
|   | Other expenses | $600'000 |
|   | Net | $64'500'000 |
|   | sold by company | 3'500'000 |
|   | sold by shareholders | 100'000 |
|   | Total shares sold | 3'600'000 |
|   | Option to underwriters | 200'000 |

| Revenues | 2009 | 2008 |
|---|---|---|
| Amount | $100'000'000 | $20'000'000 |
| Growth | 400% | |
| Number of employees | 200 | |
| Avg. val. of stock per emp. | $250'000 | |

| VCs | Round | Date | Amount | # Shares | Price | Valuation | % |
|---|---|---|---|---|---|---|---|
|   | Seed / A | Apr-03 | $1'000'000 | 10'000'000 | $0.10 | $2'500'000 | 40.0% |
|   | B | Dec-03 | $10'000'000 | 9'090'909 | $1.10 | $40'000'000 | 25.0% |
|   | C | Sep-05 | $15'000'000 | 4'098'701 | $3.66 | $150'000'000 | 10.0% |
|   | Total |   | $26'000'000 | 23'189'610 |   |   |   |

*Table i: A fictitious example of equity split in a startup at IPO time.*

# Equity allocation by type of position in the start-up

| Field | Startups # | CEO # | CEO % | VP / CXO # | VP / CXO % | CFO # | CFO % | Director # | Director % |
|---|---|---|---|---|---|---|---|---|---|
| Biotech | 114 | 48 | 2.3% | 85 | 0.7% | 55 | 0.5% | 79 | 0.3% |
| Medtech | 20 | 7 | 2.8% | 16 | 1.0% | 14 | 0.6% | 11 | 0.3% |
| Internet | 96 | 30 | 3.5% | 66 | 1.1% | 49 | 0.6% | 53 | 0.4% |
| Software | 56 | 14 | 2.9% | 41 | 1.2% | 29 | 0.6% | 33 | 0.3% |
| HW/Comp./Tel | 60 | 23 | 3.7% | 47 | 0.9% | 38 | 0.6% | 32 | 0.3% |
| Semiconductor | 35 | 14 | 2.7% | 23 | 0.8% | 19 | 0.6% | 15 | 0.5% |
| Energy/Env. | 16 | 7 | 1.9% | 13 | 0.6% | 14 | 0.5% | 9 | 0.1% |
| Other | 6 | 3 | 2.9% | 6 | 0.5% | 6 | 0.4% | 4 | 0.2% |
| Overall | 403 | 146 | 2.9% | 297 | 0.9% | 224 | 0.6% | 236 | 0.3% |

*Table ii: Data on equity split vs. fields of activity*

| Period of foundation | Startups # | CEO # | CEO % | VP / CXO # | VP / CXO % | CFO # | CFO % | Director # | Director % |
|---|---|---|---|---|---|---|---|---|---|
| 1965 | 2 | | | | | 1 | 0.6% | | |
| 1970 | 1 | | | 1 | 1.0% | | | | |
| 1975 | 5 | 2 | 5.5% | 4 | 2.5% | | | | |
| 1980 | 15 | 3 | 2.5% | 10 | 1.8% | 6 | 0.6% | | |
| 1985 | 10 | 4 | 2.3% | 9 | 1.0% | 2 | 0.5% | 4 | 0.6% |
| 1990 | 43 | 19 | 3.0% | 30 | 1.1% | 19 | 0.6% | 26 | 0.4% |
| 1995 | 98 | 36 | 3.2% | 74 | 0.9% | 59 | 0.6% | 52 | 0.4% |
| 2000 | 114 | 43 | 2.9% | 87 | 0.7% | 78 | 0.6% | 74 | 0.3% |
| 2005 | 97 | 36 | 2.6% | 71 | 0.7% | 55 | 0.5% | 72 | 0.3% |
| 2010 | 18 | 3 | 3.2% | 11 | 1.0% | 4 | 0.4% | 8 | 0.2% |
| Overall | 403 | 146 | 2.9% | 297 | 0.9% | 224 | 0.6% | 236 | 0.3% |

*Table iii: Data on equity split vs. periods of foundation*

| Geography | Startups # | CEO # | CEO % | VP / CXO # | VP / CXO % | CFO # | CFO % | Director # | Director % |
|---|---|---|---|---|---|---|---|---|---|
| Silicon Valley | 155 | 66 | 3.5% | 127 | 1.0% | 92 | 0.7% | 95 | 0.3% |
| Boston Area | 56 | 27 | 2.6% | 50 | 0.8% | 33 | 0.5% | 44 | 0.3% |
| California | 33 | 15 | 2.6% | 24 | 0.8% | 19 | 0.6% | 23 | 0.2% |
| West Coast | 17 | 8 | 2.8% | 14 | 0.9% | 8 | 0.5% | 12 | 0.2% |
| East Coast | 34 | 11 | 2.4% | 25 | 0.9% | 20 | 0.4% | 22 | 0.3% |
| Midwest | 17 | 8 | 2.4% | 16 | 0.6% | 12 | 0.4% | 11 | 0.2% |
| France | 28 | 1 | 2.1% | 17 | 1.1% | 13 | 0.9% | 8 | 0.5% |
| Switzerland | 17 | | | 3 | 0.5% | 4 | 0.6% | 6 | 0.6% |
| UK | 9 | 4 | 1.5% | 4 | 0.5% | 5 | 0.4% | 3 | 0.4% |
| Other EU | 13 | 3 | 1.5% | 7 | 1.3% | 7 | 0.5% | 4 | 0.2% |
| Israel | 4 | | | 2 | 0.4% | 2 | 0.4% | 1 | 0.4% |
| China | 12 | 1 | 1.0% | 4 | 1.2% | 4 | 1.0% | 4 | 1.3% |
| Canada | 1 | | | | | | | | |
| Other | 7 | 2 | 2.7% | 4 | 0.9% | 5 | 0.4% | 3 | 0.1% |
| Overall | 403 | 146 | 2.9% | 297 | 0.9% | 224 | 0.6% | 236 | 0.3% |

*Table iv: Data on equity split vs. geography*

# Academic Licensing

| Company | University | Year | Field | Founders | University | Series A | Total | Series A | Total | Market Cap. | Theoretical Value at Exit | Foundation | Post Series A | Post VC rounds | Exit | Note |
|---|---|---|---|---|---|---|---|---|---|---|---|---|---|---|---|---|
| Google | Stanford | 1998 | Internet | 76'980'608 | 1'842'000 | 15'360'000 | 109'626'600 | $960'000 | $41'000'000 | $24'000'000'000 | $168'000'000 | 2.3% | 2.0% | 1.0% | 0.7% | 0) |
| Akamai | MIT | 1998 | Internet | 22'081'500 | 682'110 | 20'700'000 | 37'500'000 | $8'700'000 | $113'000'000 | $2'366'000'000 | $177'734'860 | 3.0% | 1.6% | 1.8% | 0.7% | |
| Lycos | CMU | 1996 | Internet | 9'000'000 | 1'000'000 | | 14'316'000 | $1'250'000 | $1'250'000 | $230'000'000 | $16'065'940 | 10.0% | 10.0% | 8.9% | 7.0% | |
| Nanosys | Harvard | 2001 | Electronics | 2'290'000 | 160'000 | 5'500'000 | 38'000'000 | $1'650'000 | $54'000'000 | $400'000'000 | $1'200'000 | 6.5% | 2.0% | 0.4% | 0.3% | |
| Soitec | CEA LETI | 1992 | Electronics | 12'070'000 | 1'350'000 | 3'907'000 | 14'784'000 | $765'000 | $18'500'000 | $147'000'000 | $38'822'000 | 10.1% | 7.8% | 4.8% | 2.6% | |
| Gevo | Caltech | 2005 | Energy | 1'000'000 | 200'000 | 1'000'000 | 14'500'000 | $500'000 | $90'000'000 | $500'000'000 | $3'000'000 | 16.7% | 9.1% | 1.4% | 0.6% | |
| Mascoma | Darmouth | 2005 | Energy | 1'300'000 | 400'000 | 10'000'000 | 47'000'000 | $9'000'000 | $157'000'000 | $1'100'000'000 | $5'500'000 | 23.5% | 3.4% | 0.9% | 0.5% | |
| A123 | MIT | 2001 | Energy | 3'800'000 | 200'000 | 831'208 | 59'819'233 | $831'208 | $301'550'930 | $1'500'000'000 | $3'000'000 | 5.0% | 4.1% | 0.3% | 0.2% | |
| Cambridge Heart | MIT | 1990 | Medtech | 2'087'622 | 180'000 | 6'700'000 | 9'033'000 | $67'000'000 | $10'000'000 | $90'000'000 | $15'23'559 | 7.9% | 2.0% | 2.0% | 1.7% | |
| Sontra Medical | MIT | 1996 | Medtech | 7'000'000 | 558'597 | 10'057'471 | 161'557'471 | $7'000'000 | $10'400'000 | $10'000'000 | $174'562 | 7.4% | 3.2% | 3.4% | 1.7% | |
| Cubist | MIT | 1992 | Biotech | 942'043 | 74'619 | 714'286 | 5'300'000 | $500'000 | $20'000'000 | $60'000'000 | $508'766 | 7.3% | 4.3% | 1.4% | 0.8% | 1) |
| Nanogen | Salk | 1993 | Biotech | 768'336 | 40'923 | 2'340'000 | 13'683'865 | $3'510'000 | $44'000'000 | $198'000'000 | $450'153 | 5.1% | 1.3% | 0.3% | 0.2% | |
| Acusphere | MIT | 1993 | Biotech | 416'664 | 25'398 | 775'000 | 12'719'450 | $775'000 | $77'185'190 | $178'000'000 | $355'000 | 5.7% | 2.1% | 0.2% | 0.1% | |
| Genometrix | MIT | 1993 | Biotech | 8'000'000 | 950'000 | 107'000 | 16'200'000 | $1'800'000 | $17'000'000 | Did not go public | | 10.6% | 10.5% | 5.9% | 7.9% | |
| Sangamo Bio | Johns Hopkins | 1995 | Biotech | 3'900'000 | 75'000 | 791'250 | 15'149'250 | $750'000 | $16'750'000 | $106'044'750 | $525'000 | 1.9% | 1.6% | 0.5% | 0.5% | |
| Corcept | Stanford | 1996 | Biotech | 7'500'000 | 30'000 | 3'500'000 | 26'700'000 | $700'000 | $41'700'000 | $321'000'000 | $360'000 | 0.4% | 0.3% | 0.1% | 0.1% | |
| Rigel | Stanford | 1996 | Biotech | 2'400'000 | 215'000 | 7'500'000 | 43'000'000 | $6'000'000 | $36'000'000 | $302'000'000 | $1'500'000 | 8.2% | 2.1% | 0.1% | 0.1% | |
| Neurometrix | MIT | 1996 | Biotech | 2'738'155 | 1'343'000 | 875'000 | 211'164'763 | $192'500 | $42'000'000 | $92'000'000 | $44'463'552 | 32.9% | 27.1% | 6.3% | 4.9% | 2) |
| Paratek | Tufts | 1996 | Biotech | 6'500'000 | 500'000 | 1'500'000 | 26'500'000 | $1'500'000 | $130'000'000 | $575'000'000 | $63'25'000 | 7.1% | 5.9% | 1.9% | 1.1% | |
| Argos Therapeutics | Virginia | 1997 | Biotech | 19'448 | 5'192 | 56'935 | 7'000'000 | $1'900'000 | $82'000'000 | $234'000'000 | $21'273 | 21.1% | 6.4% | 0.1% | 0.0% | |
| Genomatica | UCSD | 1998 | Biotech | 2'900'000 | 350'000 | 3'500'000 | 56'000'000 | $3'500'000 | $84'000'000 | $226'000'000 | $10'54'667 | 10.8% | 5.2% | 0.6% | 0.5% | |
| Celladon | U. California | 2000 | Biotech | 24'000 | 1'744 | 45'000 | 153'000'000 | $45'00'000 | $124'000'000 | $340'000'000 | $2'624 | 6.8% | 2.5% | 0.0% | 0.0% | 3) |
| Momenta | MIT | 2001 | Biotech | 2'093'752 | 138'772 | 2'676'638 | 117'30'012 | $6'100'000 | $45'400'000 | $159'250'000 | $11'31'491 | 6.2% | 2.8% | 1.0% | 0.7% | |
| Oncomed | U. Michigan | 2001 | Biotech | 3'000'000 | 355'213 | 18'000'000 | 142'000'000 | $18'000'000 | $187'000'000 | $378'499'407 | $946'816 | 10.6% | 2.0% | 0.3% | 0.2% | 4) |
| Regado | Duke | 2001 | Biotech | 4'900'000 | 191'250 | 5'800'000 | 156'000'000 | $5'800'000 | $138'000'000 | $475'000'000 | $383'307 | 3.8% | 1.8% | 0.1% | 0.1% | |
| Tetralogic Pharma | Princeton | 2003 | Biotech | 2'864'000 | 165'000 | 8'000'000 | 153'000'000 | $8'000'000 | $66'000'000 | $400'000'000 | $242'647 | 5.4% | 1.5% | 0.1% | 0.1% | |
| Bind Therapeutics | MIT | 2006 | Biotech | 25'25'000 | 341'613 | 2'461'000 | 229'000'000 | $2'500'000 | $86'000'000 | $340'000'000 | $27'713'748 | 11.9% | 6.4% | 1.5% | 0.8% | |
| Tetraphase | Harvard | 2006 | Biotech | 4'090'000 | 910'000 | 10'400'000 | 256'000'000 | $10'400'000 | $80'000'000 | $400'000'000 | $900'000 | 18.2% | 5.9% | 0.4% | 0.2% | |
| Heat Biologics | U. Miami | 2008 | Biotech | 800'000 | 70'000 | 1'800'000 | 3'800'000 | $3'900'000 | $9'200'000 | $120'000'000 | $700'000 | 8.0% | 2.6% | 1.8% | 0.6% | |
| Verastem | Whitehead | 2010 | Biotech | 21'574'571 | 583'333 | 41'500'000 | 117'00'000 | $16'000'000 | $68'000'000 | $219'000'000 | $5'806'815 | 18.5% | 7.6% | 5.0% | 2.7% | |
| Covagen | ETH Zurich | 2006 | Biotech | 59'260 | 3'703 | 37'037 | 11'159'155 | $444'444 | $40'450'000 | $208'000'000 | $664'470 | 5.9% | 3.7% | 0.3% | 0.3% | |
| Glycovaxyn | ETH Zurich | 2004 | Biotech | 6'400 | 600 | 10'000 | 247'156 | $1'100'000 | $50'500'000 | $200'000'000 | $485'523 | 8.6% | 3.5% | 0.2% | 0.2% | |
| Molecular Partners | U. Zurich | 2004 | Biotech | 41'600'000 | 400'000 | 481'021 | 227'03'450 | $18'500'000 | $56'830'000 | $508'500'000 | $91'53'000 | 8.0% | 7.3% | 2.3% | 1.8% | |
| Speechworks | MIT | 1994 | Software | 2'800'000 | | 2'475'000 | 165'00'000 | $2'500'000 | $58'000'000 | $579'997'360 | | | | | | 5) |
| Imatron | UCSF | 1981 | Medtech | | | | | | | | | | | | | 6) |
| Kopin | MIT | 1984 | Semicon | | | | | | | | | | | | | 7) |
| Glaukos | UC Irvine | 1998 | Medtech | | | | | | | | | | | | | 8) |
| Genentech | City of Hope | 1976 | Biotech | | | | | | | | | | | | | 9) |
| | | | Average | $4'594'946 | | | | | $70'491'645 | $1'120'099'743 | $8'084'837 | 9.6% | 4.8% | 1.7% | 1.2% | |
| | | | Median | $2'500'000 | | | | | $55'415'000 | $302'000'000 | $1'093'079 | 7.9% | 3.4% | 0.9% | 0.5% | |

*Table v: Equity allocation for academic IP licenses*

0) Stanford is a little surprising with no data on company such as MIPS, Atheros, Rambus, Numerical Technologies.
1) Cubist was a 2% equity up to series B.
2) MIT only had 100'000 shares for the license, but then invested in preferred shares. There was also a 2% royalty on sales.
3) A stock split harshly diluted founders and univesity.
4) This was a 0.25% undilutable stake. The equity numbers at foundation and series A are back-computed.
5) No equity, 4% royalty up to $6M and then 1%.
6) No equity, 2% royalty on sales.
7) No equity, 3% royalty on sales.
8) no equity but $2.7M + low %
9) 2% royalty on sales

**Equity and Startups**

Startups have become in less than 50 years a major component of innovation and economic growth. An important feature of the startup phenomenon has been the wealth created through equity in startups to all stakeholders. These include the startup founders, the investors, and also the employees through the stock-option mechanism and universities through licenses of intellectual property. In the employee group, the allocation to important managers like the chief executive, vice-presidents and other officers, and independent board members is also analyzed. This report analyzes how equity was allocated in more than 400 startups, most of which had filed for an initial public offering. The author has the ambition of informing a general audience about best practice in equity split, in particular in Silicon Valley, the central place for startup innovation.

**About the author**

Hervé Lebret has been working in the startup world for more than 20 years. Since 2005, he has been in charge of support to startup creation at EPFL, the Swiss Federal Institute of Technology in Lausanne. He was before with Index Ventures, a pan-European venture capital firm which invested in Skype, mysql, Numeritech, Virata, Genmab. He used that experience to write in 2007 the book "Start-Up, what we may still learn from Silicon Valley" and the blog www.startup-book.com. Since 2010, he has also been doing research on high-tech startups with a particular focus on Silicon Valley and Stanford University. Lebret was trained in science and engineering, he is a graduate of Ecole Polytechnique (1987) and Stanford University (1990). He did his PhD in 1994 on the topic of convex optimization and its applications, which he still teaches at EPFL in addition to teaching entrepreneurship. He was a researcher in applied mathematics until he switched to venture capital in 1997.